%% file: main.tex
\documentclass[%
reprint,
superscriptaddress,
 amsmath,amssymb,
 aps,
]{revtex4-2}

\usepackage[colorlinks,citecolor=blue,linkcolor=blue,urlcolor=blue]{hyperref}
\usepackage{graphicx}% Include figure files
\usepackage{dcolumn}% Align table columns on decimal point
\usepackage{bm}% bold math
\usepackage{subcaption}
\usepackage{upgreek}
\usepackage{xcolor}
\usepackage{tabularx}
\usepackage{booktabs}
\usepackage{afterpage}
\usepackage{float}
\input macros.tex

\begin{document}

\title{Fluid-mediated sources of granular temperature at finite Reynolds numbers}

\author{Aaron M. Lattanzi}
\email[Email address for correspondence: ]{alattanz@umich.edu}
\affiliation{University of Michigan, Department of Mechanical Engineering, Ann Arbor, MI}

\author{Vahid Tavanashad}
\affiliation{Iowa State University, Department of Mechanical Engineering, Ames, IA}

\author{Shankar Subramaniam}
\affiliation{Iowa State University, Department of Mechanical Engineering, Ames, IA}

\author{Jesse Capecelatro}
\affiliation{University of Michigan, Department of Mechanical Engineering, Ann Arbor, MI}
\affiliation{University of Michigan, Department of Aerospace Engineering, Ann Arbor, MI}

\date{\today}

\begin{abstract}
We derive analytical solutions for hydrodynamic sources and sinks to granular temperature in moderately dense suspensions of elastic particles at finite Reynolds numbers. Modeling the neighbor-induced drag disturbances with a Langevin equation allows an exact solution for the joint fluctuating acceleration-velocity distribution function $P\left(v^{\prime},a^{\prime};t\right)$. Quadrant-conditioned covariance integrals of $P\left(v^{\prime},a^{\prime};t\right)$ yield the hydrodynamic source and sink that dictate the evolution of granular temperature. Analytical predictions are in agreement with benchmark data obtained from particle-resolved direct numerical simulations and show promise as a general theory from gas--solid to bubbly flows.
\end{abstract}

\maketitle
Hydrodynamic interactions between viscous fluids and a disperse phase (particles, droplets, or bubbles) give rise to complex dynamics that are crucial to many engineering and environmental applications. From the production of biofuels to post-combustion carbon capture, multiphase reactors are at the heart of nearly all chemical transformation processes. Furthermore, environmental systems such as gravity currents, debris flows, and sand dunes are also of great societal interest. Broadly speaking, the aforementioned examples are characterized by turbulent fluid flow and moderate to high solids volume fractions. The kinetic theory of rapid granular flows is now generally accepted as a valid description of moderately dense particulate systems. Namely, Enskog theory rigorously connects granular gases to continuum equations for the solids-phase moments (e.g., mass, momentum, and granular temperature) \cite{lun_kinetic_1984,garzo_dense_1999,goldhirsch_rapid_2003}. Since granular kinetic theories neglect fluid-particle interactions, they are incapable of predicting solids motion in fluidized systems. Specifically, ignoring fluid effects on granular temperature has implications on the transport coefficients. To address this deficiency, driven systems with stochastic velocity fluctuations have been theoretically \cite{puglisi_clustering_1998,cafiero_two-dimensional_2000,pagonabarraga_randomly_2001,srebro_exactly_2004} and experimentally \cite{yu_velocity_2020} examined and incorporated into Chapman--Enskog expansions \cite{garzo_enskog_2012,khalil_unified_2020}. Additionally, hydrodynamic source and sink closures have been proposed from phenomenological scaling arguments and multipole simulations \cite{koch_particle_1999}. However, closures for hydrodynamic sources and sinks to granular temperature have not been validated at finite Reynolds numbers and solids volume fraction. We show, for the first time, consistency between analytical solutions for sources and sinks to granular temperature and data obtained from particle-resolved direct numerical simulation (PR--DNS) at finite Reynolds numbers and solids volume fraction.

\begin{figure}[h]
     \centering
         \includegraphics[width=0.85\columnwidth]{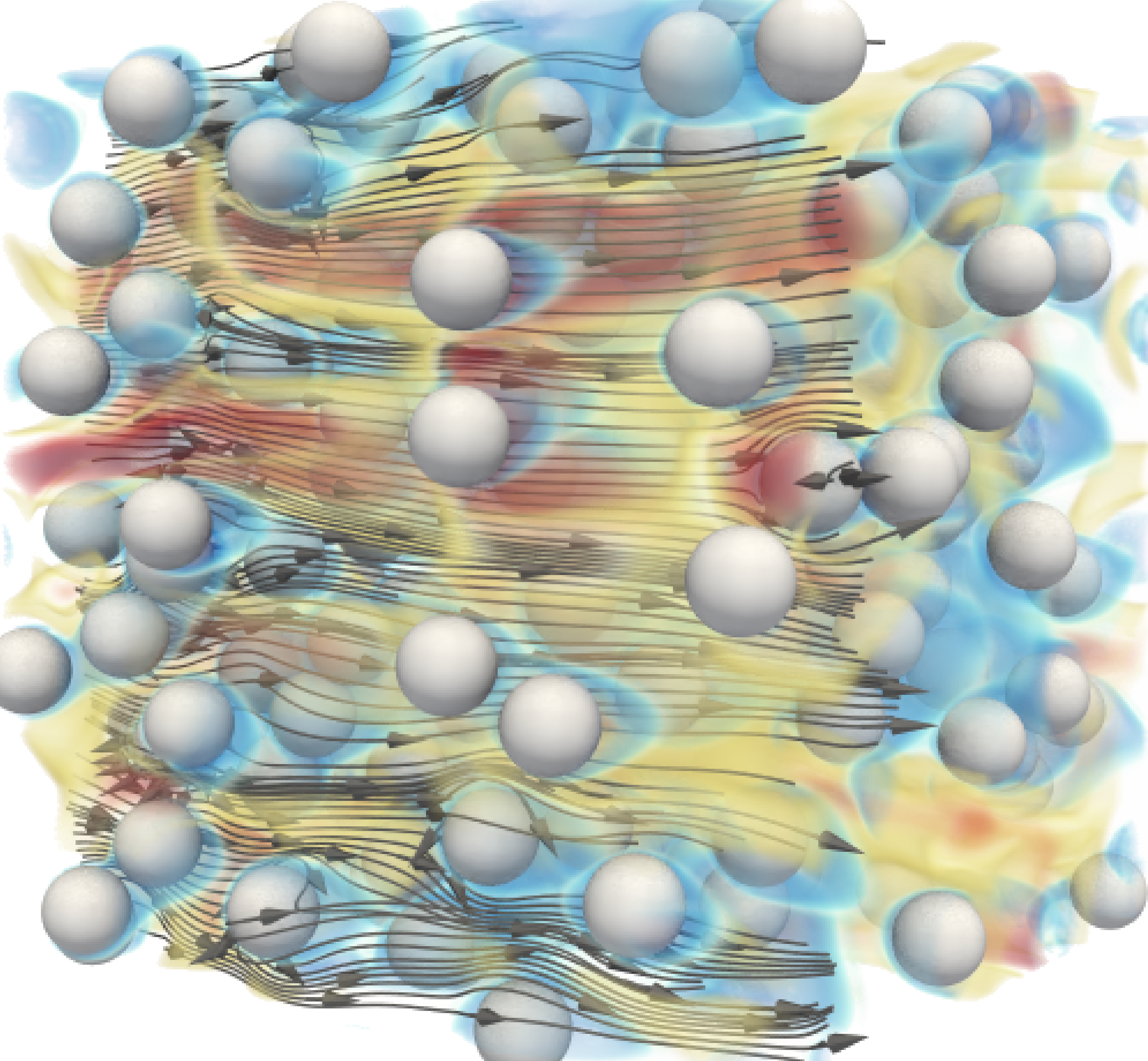}
         \caption{PR--DNS of homogeneous fluidization at ${\rm Re}_m=20$; $\rho_p/\rho_f=1000$; and $\langle \phi \rangle=0.1$. Arrows denote fluid streamlines, velocity magnitude shown in color.}
         \label{fig:PRDNS}
\end{figure}

Homogeneous fluidization of elastic smooth spheres provides a canonical flow that allows us to isolate key physics. A constant flow is exerted on the suspensions with mean Reynolds number, ${\rm Re}_m = (1-\left\langle \phi \right\rangle)\rho_fd_p\left| \left\langle \bm{w} \right\rangle\right|/\mu_f$, which drives the particles to a steady granular temperature (see Fig.~\ref{fig:PRDNS}). Here, $\langle \phi \rangle$ is the mean solids volume fraction, $\rho_f$ is the fluid density, $d_p$ is the particle diameter, $\left\langle \bm{w} \right\rangle = \left\langle \bm{u}_f \right\rangle - \left\langle \bm{v}_p \right\rangle$ is the mean slip velocity between the fluid $\left\langle \bm{u}_f \right\rangle$ and particles $\left\langle \bm{v}_p \right\rangle$, $\mu_f$ is the fluid dynamic viscosity, and $\left\langle \cdot \right\rangle$ denotes an ensemble average. The evolution of granular temperature in this system, $T=\left\langle \bm{v}_{p}^{\prime} \cdot \bm{v}_{p}^{\prime}\right\rangle/3$, is dictated by the covariance of fluctuating hydrodynamic acceleration $\bm{a}_{h}^{\prime}$ and particle velocity $\bm{v}_{p}^{\prime}$, which contains sources $S$ and sinks $\Gamma$ \cite{tenneti_stochastic_2016}
\begin{equation}
    \frac{{\rm d}T}{{\rm d}t}=\frac{2}{3}\langle \bm{a}_{h}^{\prime}\cdot\bm{v}_{p}^{\prime}\rangle=S-\Gamma, \label{eq:1}
\end{equation}
where the prime notation denotes a fluctuation from the ensemble average. Therefore, accurate descriptions for $\bm{a}_{h}^{\prime}$ are crucial to capturing $T(t)$. 

For a particle subjected to hydrodynamic forces, its total translational velocity, $\bm{v}_p = \left\langle \bm{v}_p\right \rangle + \bm{v}_p^{\prime}$, follows
\begin{equation}
    \frac{{\rm d}\bm{v}_p}{{\rm d}t}=\frac{1}{m_p}\int_{\partial \Omega} \bm{\tau}\bm{\cdot}\bm{n}\,{\rm d}S, \label{eq:2}
\end{equation}
where $m_p$ is the particle mass and surface integration of the fluid stress tensor $\bm{\tau}$ (comprised of pressure and viscous stress) gives the total hydrodynamic acceleration $\bm{a}_h$. A model for $\bm{a}_h(t)$ has been obtained in the limit of an isolated sphere and Stokes flow by \citet{basset1888treatise}, \citet{boussinesq1885application}, and \citet{oseen1927neuere} that involves a superposition of forces from the undisturbed fluid flow, quasi-steady drag, added mass, and Basset history (the so-called BBO equations). \citet{maxey_equation_1983} later extended this to non-uniform Stokes flow. At finite solids volume fractions and Reynolds numbers, analytical evaluation of the fluid stress integral is not tractable. Alternatively, correlations are often obtained from PR--DNS by ensemble averaging the net hydrodynamic force acting on a suspension. However, as shown in Fig.~\ref{fig:PRDNS}, particles interact with fluid wakes generated by their neighbors, leading to a distribution of hydrodynamic forces \cite{akiki_force_2016} that drive relative motion between particles. The application of a drag correlation, obtained from ensemble averaging, cannot reproduce the hydrodynamic force distributions obtained from PR--DNS. To properly account for said distribution, we apply the correlation of \citet{tenneti_direct_2010} to the instantaneous particle velocity and include a stochastic fluctuation $\bm{a}_h^{\prime\prime}$ 
\begin{align}
    \frac{1}{m_p}\int_{\partial \Omega} \bm{\tau}\bm{\cdot}\bm{n}\,{\rm d}S &= \frac{1}{\tau_d}  \left(\left\langle \bm{u}_f\right \rangle - \bm{v}_p \right) + \bm{a}_h^{\prime\prime} \label{eq:3}
\end{align}
where $1/\tau_d = F\left(\langle \phi \rangle,{\rm Re}_m\right) \left(1 - \langle \phi \rangle \right)/\tau_p$ is the hydrodynamic time scale, and $\tau_p=\rho_p d_p^2/(18 \mu_f)$ is the Stokes time scale. Applying $\langle \cdot \rangle$ to Eq.~\eqref{eq:3} gives $\left\langle \bm{a}_h \right \rangle$. Removal of $\left\langle \bm{a}_h \right \rangle$ yields a reference frame that moves with the mean particle velocity and $\bm{a}_h^{\prime} = -\bm{v}_p^{\prime}/\tau_d + \bm{a}_h^{\prime\prime}$.

At the conditions considered here, hydrodynamic disturbances are attributed to pseudo-turbulent kinetic energy generated by the fluid boundary layers of neighboring particles \cite{mehrabadi_pseudo-turbulent_2015,shallcross_volume-filtered_2020}. As a result, we describe $\bm{a}_h^{\prime\prime}$ with an appropriate acceleration Langevin equation \cite{lattanzi_stochastic_2021}
\begin{subequations}
\begin{align}
\mathrm{d}\bm{v}_p^{\prime}&= -\frac{1}{\tau_d} \bm{v}_p^{\prime} \, \mathrm{d}t + \bm{a}_{h}^{\prime\prime} \, \mathrm{d}t,  \label{eq:4a}\\
\mathrm{d}\bm{a}_{h}^{\prime\prime} &=  -\frac{1}{\tau_{a^{\prime\prime}}} \bm{a}_{h}^{\prime\prime} \, \mathrm{d}t + \sqrt{\frac{2}{\tau_{a^{\prime\prime}}}} \sigma_{a^{\prime\prime}} \, \mathrm{d}\bm{W}_t,
\label{eq:4b}
\end{align}
\end{subequations}
where $\tau_{a^{\prime\prime}}$ is the integral time scale of the stochastic acceleration, $\sigma_{a^{\prime\prime}}$ is the standard deviation, and $\mathrm{d}\bm{W}_t$ is a Wiener process increment. 

To quantify $S$ and $\Gamma$, one must evaluate the quadrant-conditioned acceleration-velocity covariance. Sources occur when the fluctuating acceleration is aligned with the fluctuating particle velocity (quadrants 1 and 3 in $a^{\prime}-v^{\prime}$ phase space). Analogously, sinks occur for the converse case (quadrants 2 and 4). Thus, we seek the probability distribution resulting from Eqs.~\eqref{eq:4a}--\eqref{eq:4b}. For constant coefficients, we derive the acceleration-velocity distribution in supplementary material \S~1. Here, we report the salient result that $P\left(v^{\prime},a^{\prime};t\right)=\mathcal{N}\left(\bar{\boldsymbol{\Sigma}}^{-1} \right)$ is a joint normal with
\begin{widetext}
\begin{align}
 &\bar{\boldsymbol{\Sigma}} = \left[ \arraycolsep=7pt\def\arraystretch{2.5} \begin{array}{cc}
    \frac{1}{\tau_d^2}\sigma_{v^{\prime}}^2(t) - \frac{1}{\tau_d}\sigma_{v^{\prime}a^{\prime\prime}}(t) +  \sigma_{a^{\prime\prime}}^2   &  -\frac{1}{\tau_d}\sigma_{v^{\prime}}^2(t) + \sigma_{v^{\prime}a^{\prime\prime}}(t) \\
    -\frac{1}{\tau_d}\sigma_{v^{\prime}}^2(t) + \sigma_{v^{\prime}a^{\prime\prime}}(t)    & \sigma_{v^{\prime}}^2(t)
    \end{array} \right], \label{eq:5}\\
&\sigma_{v^{\prime}}^2(t)  =  \sigma_{a^{\prime\prime}}^2 \hat{\tau}^{+} \left[ \tau_d \mathbb{E}_2  + 2 \hat{\tau}^{-} \left( \mathbb{E}_2 - \mathbb{E}_3 \right) +  \mathbb{C}_0 \tau_d \left(1 - \mathbb{E}_2  \right) - 2 \rho_0 \sqrt{\frac{\mathbb{C}_0 \tau_d}{\hat{\tau}^{+}}} \hat{\tau}^{-} \left( \mathbb{E}_2 - \mathbb{E}_3  \right) \right], \nonumber \\[1.0ex]
&\sigma_{v^{\prime}a^{\prime\prime}}(t) = \sigma_{a^{\prime\prime}}^2 \hat{\tau}^{+} \left[ \mathbb{E}_3   + \rho_0 \sqrt{\frac{\mathbb{C}_0 \tau_d}{\hat{\tau}^{+}}} \left( 1 - \mathbb{E}_3  \right)\right],   \quad \mathbb{E}_2 = \left\{ 1 - \exp \left( -\frac{2t}{\tau_d} \right) \right\},  \quad
 \mathbb{E}_3 = \left\{ 1 - \exp \left( -\frac{t}{\hat{\tau}^{+}} \right) \right\}, \nonumber
\end{align}
\end{widetext}

\noindent where $\hat{\tau}^{+} = \tau_d \tau_{a^{\prime\prime}} /(\tau_d + \tau_{a^{\prime\prime}})$, $\hat{\tau}^{-} = \tau_d \tau_{a^{\prime\prime}} /(\tau_d - \tau_{a^{\prime\prime}})$, $\rho_0 = \sigma_{v^{\prime}a^{\prime\prime}}/(\sigma_{v^{\prime}} \, \sigma_{a^{\prime\prime}})$ is the initial correlation coefficient between the fluctuating velocity and stochastic acceleration, and $\mathbb{C}_{0} \geq 0$ is a proportionality constant that specifies the initial velocity variance $\sigma_{v^{\prime},0}^2$ as a fraction of the steady velocity variance $\sigma_{v^{\prime},\infty}^2 = \sigma_{a^{\prime\prime}}^2 \hat{\tau}^{+} \tau_d$. Following \cite{stuart_kendalls_2010}, we derive relations for the quadrant-covariance of a joint-normal (see supplementary material \S~2) 
\begin{subequations}
 \label{eq:6}
\begin{align}
S(t)  = \frac{2\sqrt{\Sigma_{11} \Sigma_{22}}}{\pi} \left(\rho \arcsin{\rho} + \sqrt{1-\rho^2} +  \frac{\pi}{2} \rho \right), \label{eq:6a} \\
\Gamma(t)  = \frac{2\sqrt{\Sigma_{11} \Sigma_{22}}}{\pi} \left(\rho \arcsin{\rho} + \sqrt{1-\rho^2} -  \frac{\pi}{2} \rho \right), \label{eq:6b}
\end{align}
\end{subequations}
where $\rho(t)=\Sigma_{12}/\sqrt{\Sigma_{11} \Sigma_{22}}$ is the correlation coefficient evaluated at time $t$. 

Evaluation of Eq.~\eqref{eq:6} requires closure for the stochastic process. The acceleration time scale $\tau_{a^{\prime\prime}}$ is approximated with the mean-free time between successive collisions \cite{chapman_mathematical_1970}
\begin{equation}
\tau_{a^{\prime\prime}} \approx \tau_{\rm col} = \frac{d_p}{24 \langle \phi \rangle g_0} \sqrt{\frac{\pi}{T}},
\label{eq:7}
\end{equation}
where $g_0$ is the radial distribution function at contact \cite{MA1988191}. The standard deviation $\sigma_{a^{\prime\prime}}$ is closed with a correlation obtained from PR--DNS of static particle assemblies 
\begin{subequations}
\begin{align}
\sigma_{a^{\prime\prime}}  &= \sqrt{\frac{5}{9}} f_{\phi}^{\sigma_F} f_{\rm iso} \frac{(1-\langle \phi \rangle)\left| \langle \bm{w} \rangle \right|}{\tau_p}, \label{eq:8a} \\[1.0ex]
f_{\phi}^{\sigma_F} &= 6.52\langle \phi \rangle - 22.56 \langle \phi \rangle^2 + 49.90 \langle \phi \rangle^3, \label{eq:8b} \\
f_{\rm iso} &= \left( 1 + 0.15 {\rm Re}_m^{0.687} \right). \label{eq:8c}
\end{align}
\end{subequations}
Here, $f_{\rm iso}$ represents the drag correction for an isolated particle using the classical correlation of \citet{schiller1933fundamental}. We note that the $5/9$ factor in Eq.~\eqref{eq:8a} is employed in the collisionless theory here to account for force anisotropy. Namely, the force variance extracted from PR--DNS is observed to be $\sim 3\times$ larger in the streamwise direction than the transverse directions. To account for the variable memory time scale $\tau_{a^{\prime\prime}} = f\left(T(t)\right)$, Eq.~\eqref{eq:5} is integrated forward in time by applying the solution to a time step $\Delta t$.

\begin{figure}[H]
     \centering
         \includegraphics[width=0.85\columnwidth]{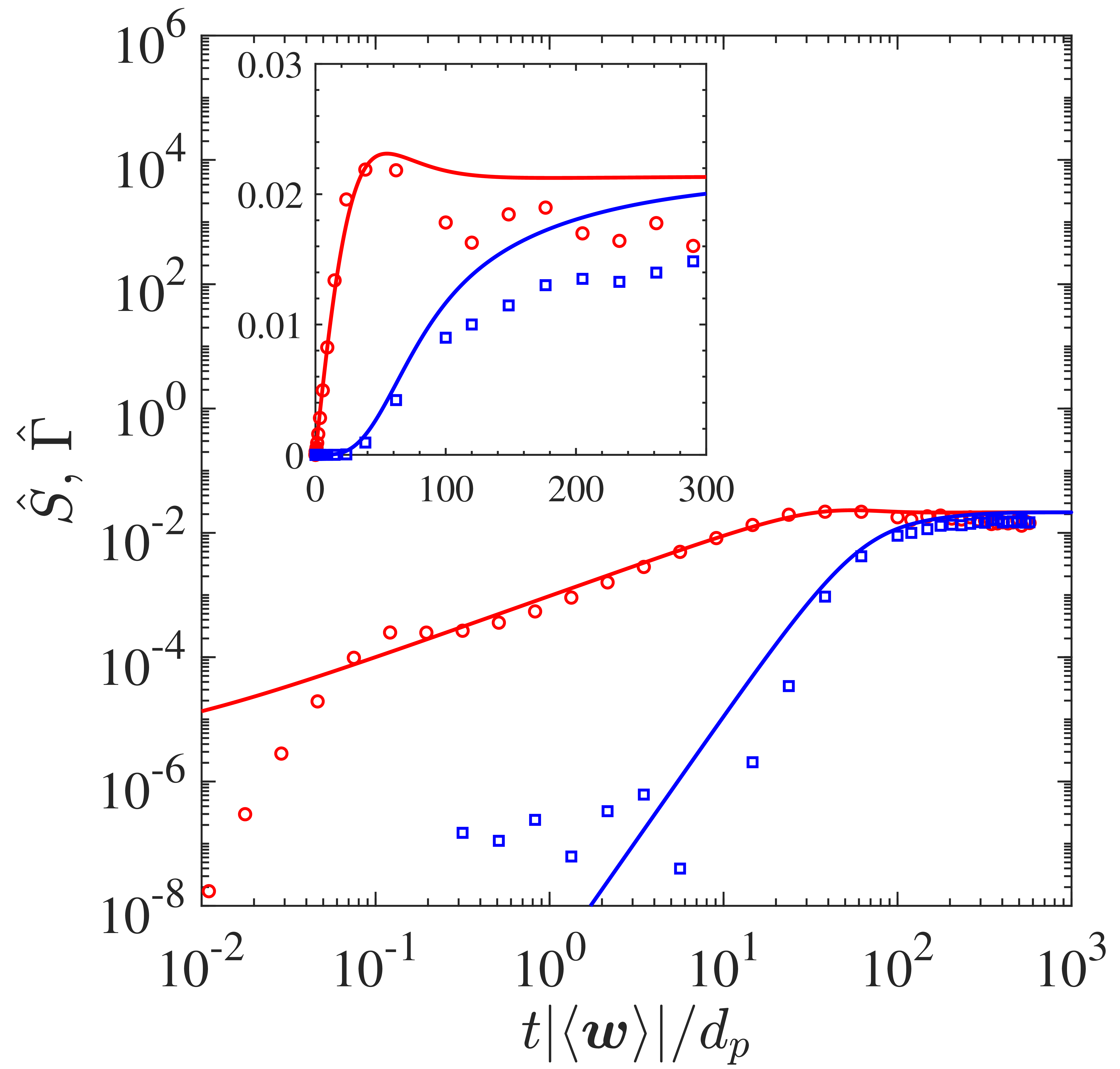}
         \caption{Non-dimensional source (red) and sink (blue) in the HHS. Analytic solution obtained from Eq.~\eqref{eq:6} with $\rho_0=1$ (lines), PR--DNS data (symbols). The inset shows the same data with linear scaling.}
         \label{fig:1}
\end{figure}

\begin{figure}[H]
     \centering
         \includegraphics[width=0.85\columnwidth]{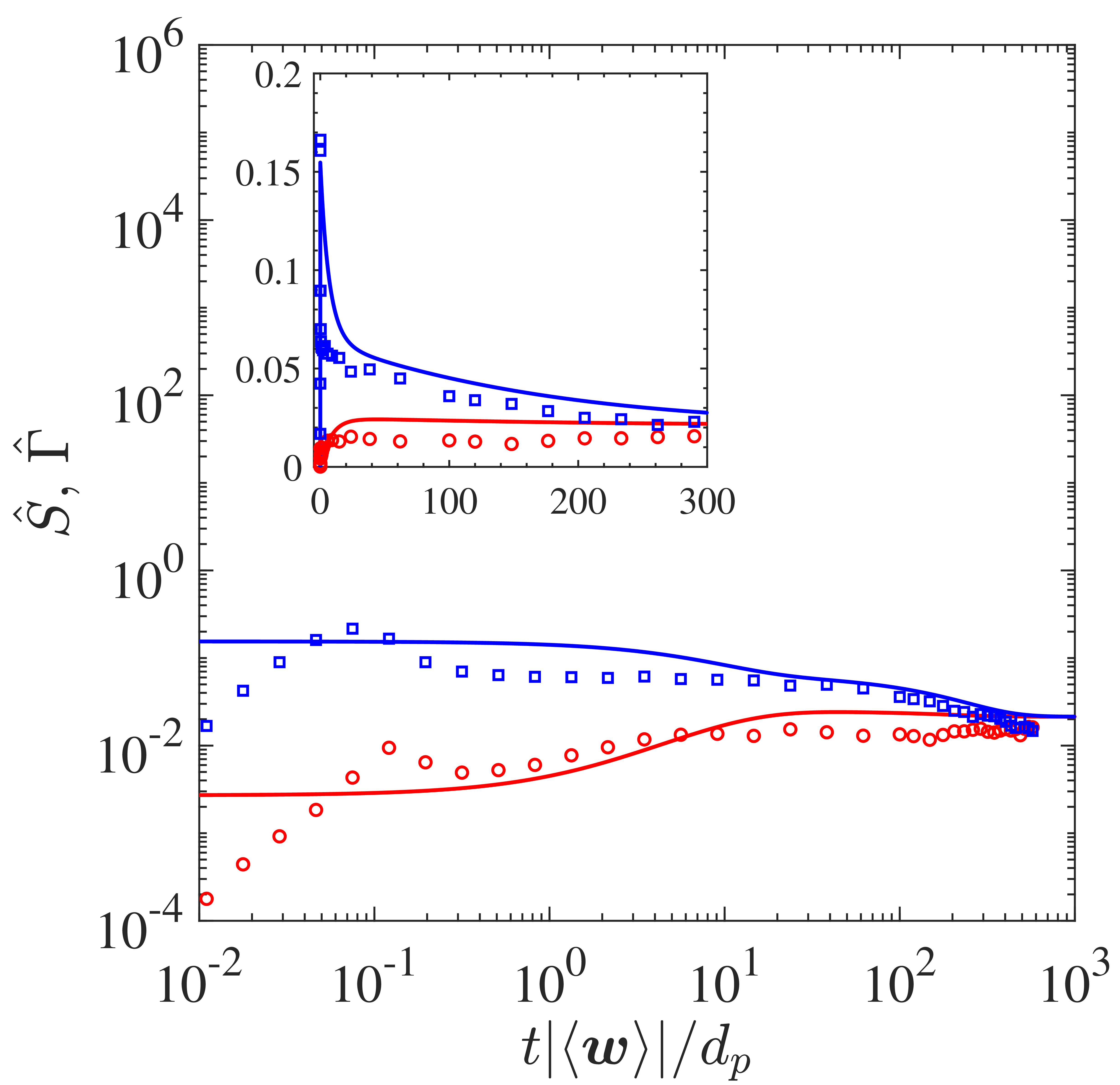}
         \caption{Same legend as Fig.~\ref{fig:1} for the HCS but with $\rho_0=-0.75$.}
         \label{fig:2}
\end{figure}

Analytical results are compared with PR--DNS benchmark data for two canonical flows at ${\rm Re}_m=20$; $\rho_p/\rho_f=1000$; $\langle \phi \rangle=0.1$. Namely, homogeneous heating (HHS) and homogeneous cooling (HCS) systems are examined in detail. In the former system, particles are initialized as $T_0 = 0$, while the latter is initialized as $T_{0} > T_{\infty}$, with $T_{\infty}$ the steady granular temperature. Considering the dimensionless source $\hat{S}(t) = S(t) \tau_p/\left((1-\langle \phi \rangle)^2 \left| \left\langle \bm{w} \right\rangle\right|^2 \right)$, and analogous dimensionless sink $\hat{\Gamma}(t)$, we observe strong agreement between analytical predictions and PR--DNS data (see Figs.~\ref{fig:1}--\ref{fig:2}). 

\begin{figure}[H]
     \centering
         \includegraphics[width=0.85\columnwidth]{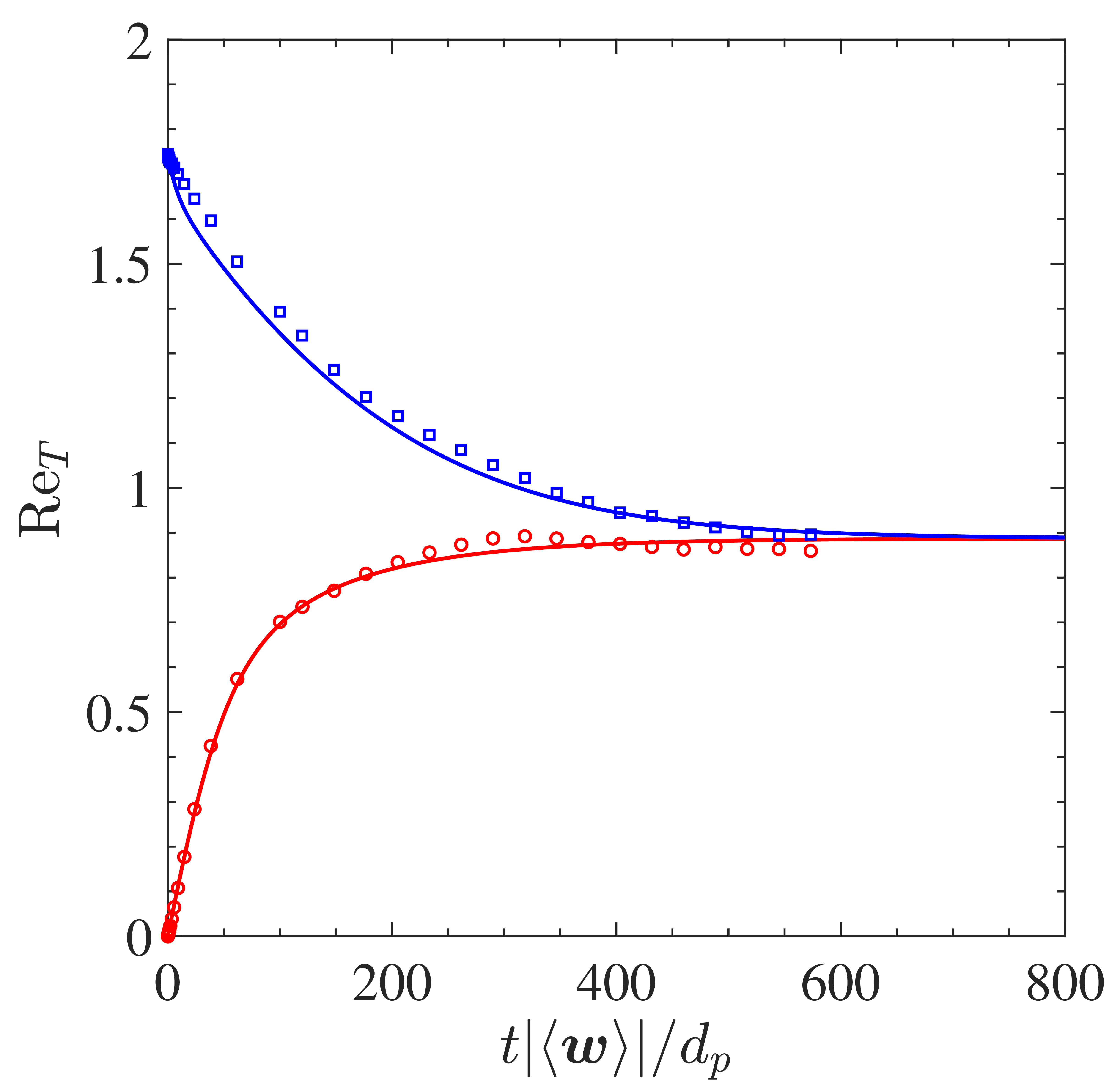}
         \caption{Evolution of non-dimensional granular temperature for the HHS (red) and HCS (blue). Analytic solution (lines), PR--DNS (symbols).}
         \label{fig:3}
\end{figure}

\begin{figure}[H]
\centering
    \includegraphics[width=0.99\columnwidth]{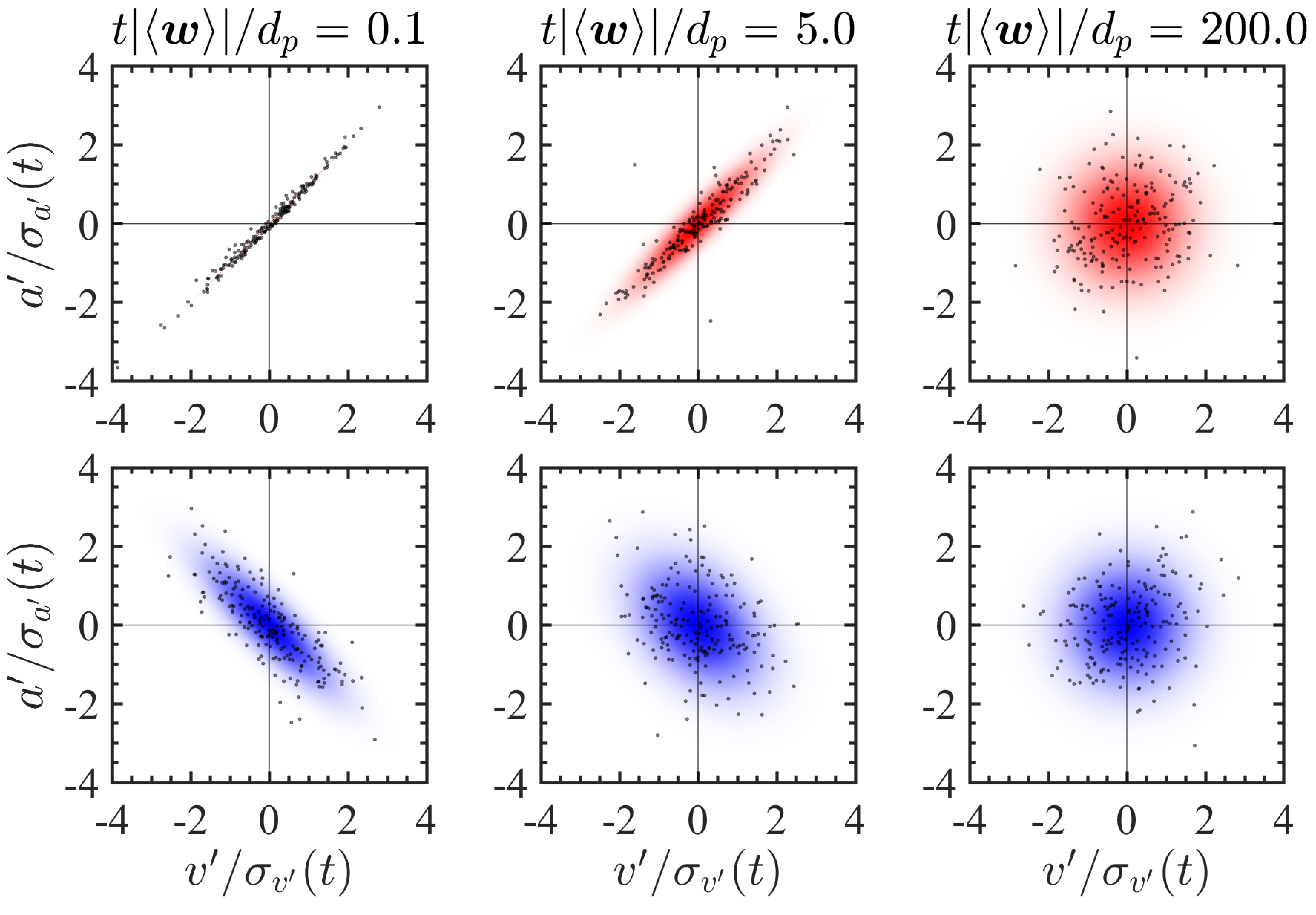}
     \caption{Joint PDFs of the fluctuating particle acceleration and fluctuating particle velocity for the HHS (top) and HCS (bottom). Analytic solution (color), PR--DNS (symbols). The first and third quadrant correspond to sources of granular temperature and the second and fourth quadrant correspond to dissipation.}
    \label{fig:4}
\end{figure}

A direct consequence of accurately characterizing $S(t)$ and $\Gamma(t)$ is that the theory captures the temporal evolution of ${\rm Re}_T = \rho_f d_p \sqrt{T}/\mu_f$ (see Fig.~\ref{fig:3}). The ability of Eq.~\eqref{eq:5} to capture the granular temperature dynamics speaks to the statistical equivalence between the acceleration Langevin model and homogeneous fluidization of elastic particles at finite Reynolds numbers. To demonstrate equivalence in a more quantitative manner we compare the acceleration-velocity probability distribution to scatter plots obtained 
from PR--DNS (see Fig.~\ref{fig:4}). Again, data extracted from PR--DNS is well characterized by $P\left(v^{\prime},a^{\prime};t\right)$ obtained from integration of Eq.~\eqref{eq:5}. In the HHS (top panels of Fig.~\ref{fig:4}), granular temperature is dominated by hydrodynamic sources (quadrants 1 and 3) at early time that become balanced by the sinks, leading to a sustained particle velocity variance. By contrast, hydrodynamic sinks (quadrants 2 and 4) are dominant in the HCS (bottom panels of Fig.~\ref{fig:4}) since the system is initialized with an over-prescribed velocity variance.


While the present theory is designed around inertial particles at high density ratios $\rho_p / \rho_f \gg1$, it is beneficial to consider the behavior across density ratio. Previous PR--DNS studies concluded that $T_{\infty} \sim \rho_f/\rho_p$ for $\rho_p/\rho_f \geq 100$ \cite{tenneti_stochastic_2016,tang_direct_2016}. However, such scaling leads to unrealistic granular temperature for low density particles. \citet{tavanashad_effect_2019} showed that $T_{\infty}$ levels off at low density ratio due to the additional dissipation associated with unsteady hydrodynamic forces (e.g., added mass and Basset history). Taking $\lim_{t\to\infty} \sigma_{v^{\prime}}^2(t)$ yields an algebraic relation $T_{\infty} = \sigma_{a^{\prime\prime}}^2 \tau_d^2 \tau_{a^{\prime\prime}}(T_{\infty})/(\tau_d + \tau_{a^{\prime\prime}} (T_{\infty}) )$.


\begin{figure}[H]
     \centering
         \includegraphics[width=0.85\columnwidth]{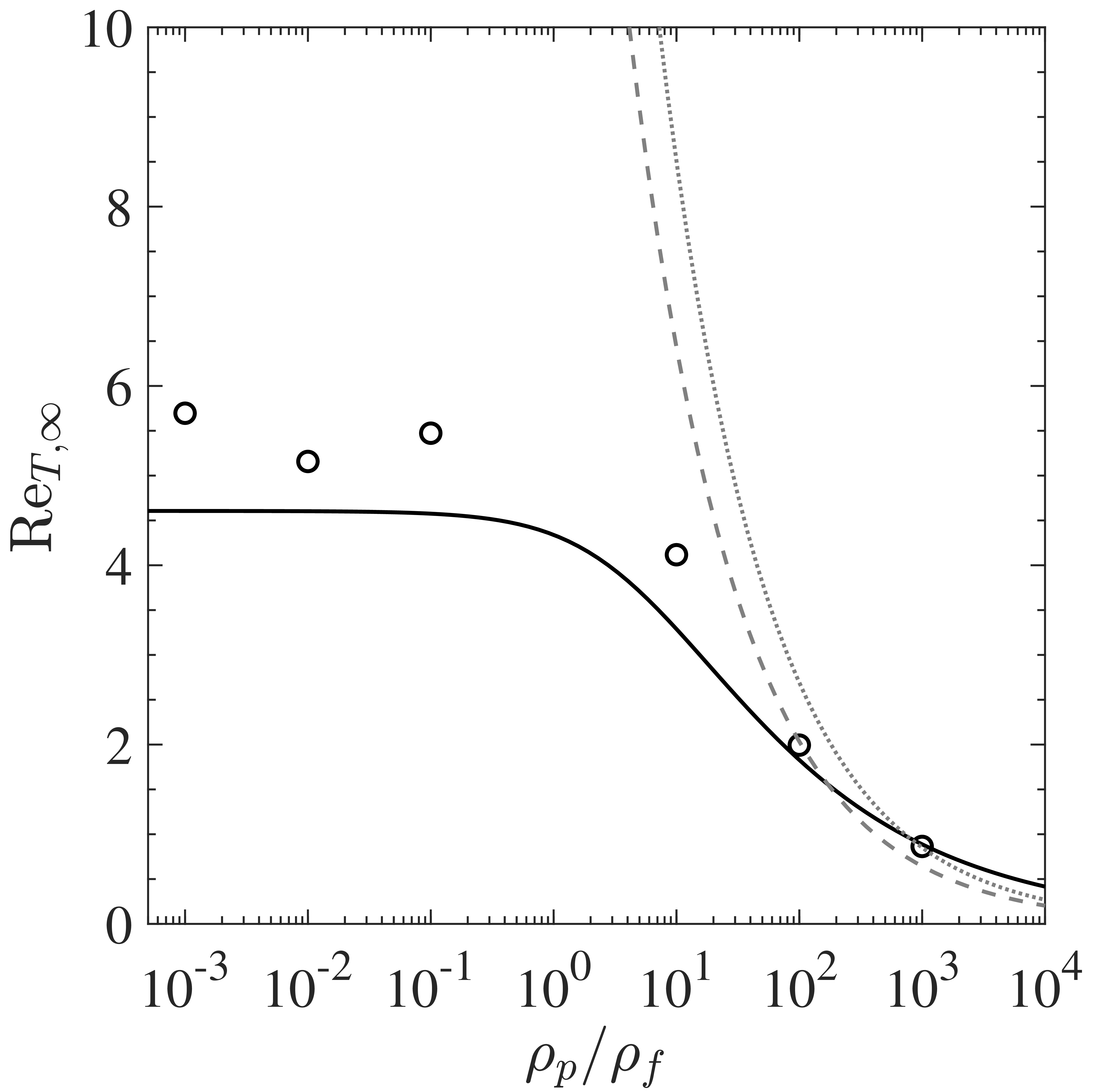}
         \caption{Non-dimensional granular temperature at steady state as a function of density ratio. Analytical solution (solid line), PR--DNS of \citet{tavanashad_effect_2019} (symbols), correlation of \citet{tenneti_stochastic_2016} (dashed line), correlation of \citet{tang_direct_2016} (dotted line).}
         \label{fig:5}
\end{figure}

\noindent We solve said relation to ascertain the behavior of the present theory across a wide range of density ratios (see Fig.~\ref{fig:5}). As opposed to earlier correlations obtained from PR--DNS of inertial particles~\cite{tenneti_stochastic_2016,tang_direct_2016}, the acceleration Langevin solution is capable of \emph{predicting} the tail off of granular temperature reported by \citet{tavanashad_effect_2019}. The asymptotic behavior of the theory may be seen by noting that $\tau_d=f(\rho_p)$ and $\tau_d \ll \tau_{a^{\prime\prime}}$ as $\rho_p/\rho_f \rightarrow 0$.

While results presented here at low density ratio are incipient, the modeling paradigm shows promise as a viable starting place for developing a theory across particle density ratio. In this spirit, $\sigma_{a^{\prime\prime}}$ correlations may be obtained beyond the static particle limit and utilized alongside drag correlations that are valid across density ratio \cite{tavana_IJMF_2021}. These efforts may allow extension of the present theory beyond inertial particles --- i.e., a general description for fluid-mediated sources to granular temperature that is valid from gas--solid to bubbly flows appears possible. 

This work was supported by National Science Foundation under grants CBET-1904742 and CBET-1438143.

\bibliography{main}

%%%%%%%%%% Merge with supplemental materials %%%%%%%%%%
\pagebreak
\widetext
\begin{center}
\textbf{\large Supplemental Material:\\ Fluid-mediated sources of granular temperature at finite Reynolds numbers}
\end{center}
%%%%%%%%%% Merge with supplemental materials %%%%%%%%%%
%%%%%%%%%% Prefix a "S" to all equations, figures, tables and reset the counter %%%%%%%%%%
\setcounter{equation}{0}
\setcounter{figure}{0}
\setcounter{table}{0}
\setcounter{page}{1}
\makeatletter
\renewcommand{\theequation}{S\arabic{equation}}
\renewcommand{\thefigure}{S\arabic{figure}}
\renewcommand{\bibnumfmt}[1]{[S#1]}
\renewcommand{\citenumfont}[1]{S#1}
%%%%%%%%%% Prefix a "S" to all equations, figures, tables and reset the counter %%%%%%%%%%

\section{The joint acceleration-velocity distribution}\label{sec:javd}
\subsection{General formulation}\label{subsec:gf}
Following our previous works on neighbor-induced drag disturbances \cite{lattanzi_stochastic_2020,lattanzi_stochastic_2021}, we consider a stochastic approach for the hydrodynamic force perturbation induced by particles in close proximity. Specifically, the following stochastic differential equations (SDEs) are considered
\begin{subequations}
\begin{align}
\mathrm{d}\bm{x}_{p}^{\prime} &= \bm{v}_{p}^{\prime} \,  \mathrm{d}t,  \label{eq:ufsde1}\\
\mathrm{d}\bm{v}_{p}^{\prime} &= -\frac{1}{\tau_d} \bm{v}_{p}^{\prime} \,  \mathrm{d}t + \bm{a}_{h}^{\prime \prime} \,  \mathrm{d}t,  \label{eq:ufsde2}\\
\mathrm{d}\bm{a}_{h}^{\prime \prime} &=  -\frac{1}{\tau_{a^{\prime\prime}}} \bm{a}_{h}^{\prime \prime} \,  \mathrm{d}t + \sqrt{\frac{2}{\tau_{a^{\prime\prime}}}} \sigma_{a^{\prime\prime}} \,  \mathrm{d}\bm{W}_t,
\label{eq:ufsde3}
\end{align}
\end{subequations}
where $\bm{x}_{p}^{\prime}$, $\bm{v}_{p}^{\prime}$, $\bm{a}_{h}^{\prime \prime}$ are the fluctuating position, velocity, and hydrodynamic acceleration, respectively, $\tau_{a^{\prime\prime}}$ is the fluctuating acceleration memory, $\sigma_{a^{\prime\prime}}=\sigma_{F^{\prime\prime}}/m_p$ is the fluctuating acceleration standard deviation, $\sigma_{F^{\prime\prime}}$ is the standard deviation in drag force, and $m_p$ is the particle mass. $\tau_d$ is the drag response time scale obtained from applying the correlation of \citet{tenneti_drag_2011} in the same manner as Appendix C of \citet{lattanzi_stochastic_2020}. When deriving the weak (distribution) solution to the set of stochastic differential equations, we consider the coefficients $(\tau_d; \, \tau_{a^{\prime\prime}}; \, \sigma_{a^{\prime\prime}})$ to be constant. Physical arguments are provided in \S~\ref{subsec:sol} for why $\sigma_{a^{\prime\prime}}$ may be approximated as constant in the homogeneous system. However, the acceleration memory is approximated with the mean-free-time between collisions (Eq. 7 in main text) and will lead to $\tau_{a^{\prime\prime}} = f(T(t))$, where $T$ is the granular temperature (velocity variance). The constant coefficient solution obtained below (see Eq.~\eqref{eq:covmat}) may be integrated forward in time to account for the functional dependence of $\tau_{a^{\prime\prime}}$. Supplementary code is provided to demonstrate how solutions derived herein yield the reported results. More specifically, we include a library of solutions ({\tt AL\_Lib.m}) that is utilized by integration routines for the homogeneous heating ({\tt AL\_HHS\_Example.m}) and homogeneous cooling ({\tt AL\_HCS\_Example.m}) cases to reproduce Figures 2--4 in the main text. 

The Fokker-Planck equation resulting from Eqs.~\eqref{eq:ufsde1}--\eqref{eq:ufsde3} is given by
\begin{equation}
\diff{P}{t} + \nabla_{\bm{x}^{\prime}} \cdot  \left(\bm{v}^{\prime}P\right) + \nabla_{\bm{v}^{\prime}} \cdot \left[\left(\bm{a}^{\prime\prime} - \frac{1}{\tau_d}\bm{v}^{\prime}\right)P\right] - \frac{1}{\tau_{a^{\prime\prime}}} \nabla_{\bm{a}^{\prime\prime}} \cdot \left(\bm{a}^{\prime\prime} P\right) = \frac{\sigma_{a^{\prime\prime}}^2}{ \tau_{a^{\prime\prime}}} \Delta_{\bm{a}^{\prime\prime}} P,
\label{eq:fp}
\end{equation}
where $P(\bm{x}^{\prime},\bm{v}^{\prime},\bm{a}^{\prime\prime};t|\bm{y}^{\prime},\bm{u}^{\prime},\bm{b}^{\prime\prime};s)$ is the probability density function and subscripts denote the phase space variables being differentiated. Considering Eq.~\eqref{eq:fp} in 1D, we integrate over $x^{\prime}$ to obtain the evolution of the joint acceleration-velocity distribution $P(v^{\prime},a^{\prime\prime};t|u^{\prime},b^{\prime\prime};s)$
\begin{equation}
\diff{P}{t} + \diff{}{v^{\prime}} \left[\left(a^{\prime\prime} - \frac{1}{\tau_d}v^{\prime}\right)P\right] - \frac{1}{\tau_{a^{\prime\prime}}} \diff{}{a^{\prime\prime}} \left(a^{\prime\prime} P\right) = \frac{\sigma_{a^{\prime\prime}}^2}{\tau_{a^{\prime\prime}}} \diffn{P}{a^{\prime\prime}}{2}. \label{eq:fp1d}
\end{equation}
Applying Fourier transform to Eq.~\eqref{eq:fp1d}, $\mathcal{F}\left[ P(\bm{x};t) \right] = \mathcal{P}(\bm{k};t)$, we obtain
\begin{equation}
\diff{\mathcal{P}}{t} + \frac{k_{v^{\prime}}}{\tau_d}\diff{\mathcal{P}}{k_{v^{\prime}}} + \left[ \frac{k_{a^{\prime\prime}}}{\tau_{a^{\prime\prime}}} - k_{v^{\prime}} \right] \diff{\mathcal{P}}{k_{a^{\prime\prime}}}  = -\frac{k_{a^{\prime\prime}}^2 \sigma_{a^{\prime\prime}}^2}{ \tau_{a^{\prime\prime}}} \mathcal{P}. \label{eq:fp1df}
\end{equation}
Employing method of characteristics (MOC), we seek a parameterizaton variable $s$
\begin{equation}
\odiff{\mathcal{P}(t(s),k_{v^{\prime}}(s),k_{a^{\prime\prime}}(s))}{s} = \diff{\mathcal{P}}{t}\odiff{t}{s} + \diff{\mathcal{P}}{k_{v^{\prime}}}\odiff{k_{v^{\prime}}}{s} + \diff{\mathcal{P}}{k_{a^{\prime\prime}}}\odiff{k_{a^{\prime\prime}}}{s}, \label{eq:mocparam}
\end{equation}
and match the coefficients of Eq.~\eqref{eq:mocparam} with Eq.~\eqref{eq:fp1df} to yield a system of equations
\begin{subequations}
\begin{alignat}{2}
&\odiff{t}{s}     &&= 1 \\[1.0ex]
&\odiff{k_{v^{\prime}}}{s} &&=  \frac{k_{v^{\prime}}}{\tau_d} \\[1.0ex]
&\odiff{k_{a^{\prime\prime}}}{s} &&=  \frac{k_{a^{\prime\prime}}}{\tau_{a^{\prime\prime}}} - k_{v^{\prime}}  \\[1.0ex]
&\odiff{\mathcal{P}}{s} &&= -\frac{k_{a^{\prime\prime}}^2 \sigma_{a^{\prime\prime}}^2}{ \tau_{a^{\prime\prime}}} \mathcal{P}. \label{eq:pode}
\end{alignat}
\end{subequations}
Integrating the system of equations gives
\begin{subequations}
\begin{alignat}{2}
&t     &&= s + C_0,  \\[1.0ex]
&k_{v^{\prime}} &&= C_1 \exp\left(\frac{s}{\tau_d}\right), \\[1.0ex] 
&k_{a^{\prime\prime}} &&= C_1 \hat{\tau}^{-} \exp \left( \frac{s}{\tau_d} \right) + C_2 \exp \left( \frac{s}{\tau_{a^{\prime\prime}}} \right), \label{eq:ka} \\[1.0ex]
&\mathcal{P} &&= C_3 \exp \left(-\frac{\sigma_{a^{\prime\prime}}^2}{2 }\left[  \left(C_1 \hat{\tau}^{-}\right)^2 \frac{\tau_d}{\tau_{a^{\prime\prime}}} \exp\left(\frac{2s}{\tau_d}\right) +  C_2^2  \exp \left( \frac{2s}{\tau_{a^{\prime\prime}}} \right) + 4 C_1 C_2 \frac{\hat{\tau}^{-} \hat{\tau}^{+}}{\tau_{a^{\prime\prime}}} \exp \left( \frac{s}{\hat{\tau}^{+}} \right) \right]\right), \label{eq:mocsol}
\end{alignat}
\end{subequations}
where $\hat{\tau}^{-} = \tau_d \tau_{a^{\prime\prime}} /(\tau_d - \tau_{a^{\prime\prime}})$ and $\hat{\tau}^{+} = \tau_d \tau_{a^{\prime\prime}} /(\tau_d + \tau_{a^{\prime\prime}})$. Taking $C_0 = 0$, we show the MOC solution is invertible, viz.
\begin{subequations}
\begin{alignat}{2}
&s     &&= t \label{eq:invert}, \\[1.0ex]
&C_1 &&= k_{v^{\prime}} \exp\left(-\frac{t}{\tau_d}\right), \\[1.0ex] 
&C_2 &&= \left[ k_{a^{\prime\prime}} - k_{v^{\prime}} \hat{\tau}^{-} \right] \exp \left( -\frac{t}{\tau_{a^{\prime\prime}}} \right). \label{eq:invert3}
\end{alignat}
\end{subequations}
To obtain a closed-form solution one must impose an initial condition $\mathcal{P}(0)$, invert constants via Eqs.~\eqref{eq:invert}--\eqref{eq:invert3}, and apply an inverse Fourier transform $\mathcal{F}^{-1}\left[ \mathcal{P}(\bm{k};t)\right] = P(\bm{x};t)$. We consider a general initial condition in the following subsection.

\subsection{Arbitrarily correlated initial condition }\label{subsec:sol}
We consider particles with specified initial velocity variance $\sigma_{v^{\prime},0}^2$ and initial acceleration-velocity covariance $\sigma_{v^{\prime}a^{\prime\prime},0} = \rho_0 \sigma_{a^{\prime\prime}} \sigma_{v^{\prime},0}$. The initial condition allows us to consider general fluidization, which will encompasses the cooling and heating systems probed by PR--DNS studies \cite{tenneti_direct_2010,tenneti_stochastic_2016,tavanashad_effect_2019}. For inertial particles, fluid boundary layers will develop on much smaller time scales than the particle velocity variance; thus, we consider the fluctuating hydrodynamic acceleration to be fully-developed (steady solution to Ornstein-Uhlenbeck process in Eq.~\eqref{eq:ufsde3}). The correlated initial condition yields
\begin{equation}
\mathcal{F} \left[ \mathcal{N}\left(  \bar{\bm{\Sigma}}^{-1} \right) \right] =  \exp\left(-\frac{ k_{a^{\prime\prime}}^2  \sigma_{a^{\prime\prime}}^2}{2} -\frac{ k_{v^{\prime}}^2  \sigma_{v^{\prime},0}^2 }{2}  - k_{a^{\prime\prime}} k_{v^{\prime}} \sigma_{v^{\prime}a^{\prime\prime},0} \right), \label{eq:initdist}
\end{equation}
where $\mathcal{N}\left(  \bar{\bm{\Sigma}}^{-1} \right)$ is the normal distribution with covariance matrix $ \bar{\bm{\Sigma}}$. Evaluating Eq.~\eqref{eq:initdist} at $k_i(0)$ gives the initial condition
\begin{equation}
\mathcal{P}(0) = \exp\left(-\frac{\sigma_{a^{\prime\prime}}^2}{2}\left[ \left(C_1 \hat{\tau}^{-}\right)^2 + C_2^2 +  2 C_1 C_2 \hat{\tau}^{-} \right] - \frac{\sigma_{v^{\prime},0}^2}{2}C_1^2 - \sigma_{v^{\prime}a^{\prime\prime},0} \left[C_1^2 \hat{\tau}^{-} + C_1 C_2 \right] \right), \label{eq:initcond}
\end{equation}
which allows the $C_3$ constant to be obtained
\begin{align}
C_3 = &\exp \left( \frac{\sigma_{a^{\prime\prime}}^2}{2}\left[ \left(C_1 \hat{\tau}^{-}\right)^2 \left(\frac{\tau_d}{\tau_{a^{\prime\prime}}} - 1\right) + 4C_1 C_2 \frac{\hat{\tau}^{-} \hat{\tau}^{+}}{\tau_{a^{\prime\prime}}} - 2C_1C_2\hat{\tau}^{-}  \right] \right) \times \label{eq:c3} \\
&  \exp \left( - \frac{\sigma_{v^{\prime},0}^2}{2}C_1^2 -  \sigma_{v^{\prime}a^{\prime\prime},0} \left[C_1^2 \hat{\tau}^{-} + C_1 C_2 \right]  \right). \nonumber
\end{align}
After inverting the constants ($C_1$; $C_2$), and a considerable amount of algebra, the joint distribution solution in wave-space is obtained
\begin{subequations}
\begin{align}
&\mathcal{P} (k_{v^{\prime}},k_{a^{\prime\prime}};t) = \exp \left( -\frac{k_{a^{\prime\prime}}^2}{2}\sigma_{a^{\prime\prime}}^2  - \frac{k_{v^{\prime}}^2}{2} \sigma_{v^{\prime}}^2(t) - k_{a^{\prime\prime}} k_{v^{\prime}} \sigma_{v^{\prime}a^{\prime\prime}}(t)  \right), \label{eq:solution} \\[1.0ex]
&\sigma_{v^{\prime}}^2(t)  =  \sigma_{a^{\prime\prime}}^2 \hat{\tau}^{+} \left[ \tau_d \mathbb{E}_2  + 2 \hat{\tau}^{-} \left( \mathbb{E}_2 - \mathbb{E}_3 \right) +  \mathbb{C}_0 \tau_d \left(1 - \mathbb{E}_2  \right) - 2 \rho_0 \sqrt{\frac{\mathbb{C}_0 \tau_d}{\hat{\tau}^{+}}} \hat{\tau}^{-} \left( \mathbb{E}_2 - \mathbb{E}_3  \right) \right], \label{eq:uutot}\\[1.0ex]
&\sigma_{v^{\prime}a^{\prime\prime}}(t) = \sigma_{a^{\prime\prime}}^2 \hat{\tau}^{+} \left[ \mathbb{E}_3   + \rho_0 \sqrt{\frac{\mathbb{C}_0 \tau_d}{\hat{\tau}^{+}}} \left( 1 - \mathbb{E}_3  \right)\right], \label{eq:au} \\[1.0ex]
&\mathbb{E}_2 = \left\{ 1 - \exp \left( -\frac{2t}{\tau_d} \right) \right\}, \\[1.0ex]
&\mathbb{E}_3 = \left\{ 1 - \exp \left( -\frac{t}{\hat{\tau}^{+}} \right) \right\}.
\end{align}
\end{subequations}
$\mathbb{C}_{0} \geq 0$ is a proportionality constant that specifies the initial velocity variance $\sigma_{v^{\prime},0}^2$ as a fraction of the steady velocity variance $\sigma_{v^{\prime},\infty}^2 = \sigma_{a^{\prime\prime}}^2 \hat{\tau}^{+} \tau_d$. Similarly, $\rho_0 \in \left[-1 \, 1 \right]$ is the initial correlation coefficient that specifies the initial covariance $\sigma_{v^{\prime}a^{\prime\prime},0}$. Since Eq.~\eqref{eq:solution} is joint-normal in wave-space, it will also be joint-normal in physical space. Furthermore, taking $\lim \mathbb{C}_0 \rightarrow 0$ and $\lim \rho_0 \rightarrow 0$ shows that the velocity variance $\sigma_{v^{\prime}}^2(t)$  and acceleration-velocity covariance $\sigma_{v^{\prime}a^{\prime\prime}}(t)$ match the solutions provided in \cite{lattanzi_stochastic_2020}.

\subsection{Total acceleration distribution}\label{subsec:tad}
The total fluctuating hydrodynamic acceleration $\bm{a}_{h}^{\prime}$, right hand side of Eq.~\eqref{eq:ufsde2}, includes contributions from deterministic drag and stochastic fluctuations
\begin{equation}
\bm{a}_{h}^{\prime} = -\frac{1}{\tau_d} \bm{v}_{p}^{\prime} + \bm{a}_{h}^{\prime\prime}.
\end{equation}
To compare with PR--DNS, we require $P(\bm{v}^{\prime},\bm{a}^{\prime};t)$ rather than $P(\bm{v}^{\prime},\bm{a}^{\prime\prime};t)$. Thus, we seek to characterize the variances and covariances of total fluctuating hydrodynamic acceleration. The covariance tensors are given by
\begin{subequations}
\begin{align}
\left\langle \bm{a}_{h}^{\prime} \otimes \bm{a}_{h}^{\prime} \right\rangle &= \frac{1}{\tau_d^2} \left\langle \bm{v}_{p}^{\prime}  \otimes  \bm{v}_{p}^{\prime} \right\rangle - \frac{1}{\tau_d} \left\langle \bm{v}_{p}^{\prime}  \otimes \bm{a}_{h}^{\prime\prime} \right\rangle + \left\langle \bm{a}_{h}^{\prime\prime} \otimes \bm{a}_{h}^{\prime\prime} \right\rangle, \label{eq:atotal} \\
\left\langle \bm{v}_{p}^{\prime} \otimes \bm{a}_{h}^{\prime} \right\rangle &= -\frac{1}{\tau_d} \left\langle \bm{v}_{p}^{\prime}  \otimes  \bm{v}_{p}^{\prime} \right\rangle + \left\langle \bm{v}_{p}^{\prime}  \otimes \bm{a}_{h}^{\prime\prime} \right\rangle. 
\end{align}
\end{subequations}
Substituting results from Eqs.~\eqref{eq:uutot}--\eqref{eq:au}, we then obtain
\begin{subequations}
\begin{align}
\sigma_{a^{\prime}}^2 &= \frac{1}{\tau_d^2}\sigma_{v^{\prime}}^2 - \frac{1}{\tau_d}\sigma_{v^{\prime}a^{\prime\prime}} +  \sigma_{a^{\prime\prime}}^2, \label{eq:aatot}\\
\sigma_{v^{\prime} a^{\prime}} &= -\frac{1}{\tau_d}\sigma_{v^{\prime}}^2 + \sigma_{v^{\prime}a^{\prime\prime}} \label{eq:autot}.
\end{align}
\end{subequations}
Therefore, the fluctuating acceleration-velocity distribution $P(\bm{v}^{\prime},\bm{a}^{\prime};t)$ is a normal distribution $\mathcal{N}\left(\bar{\boldsymbol{\Sigma}}^{-1} \right)$ with time dependent covariance tensor given by
\begin{align}
    \bar{\boldsymbol{\Sigma}} &= \left[ \begin{array}{cc}
    \sigma_{a^{\prime}}^2(t)            &  \sigma_{v^{\prime} a^{\prime}}(t)\\
    \sigma_{v^{\prime} a^{\prime}}(t)   & \sigma_{v^{\prime}}^2(t)
    \end{array} \right]. \label{eq:covmat}
\end{align}

\section{Evolution of sources \& sinks}\label{sec:ess}
\subsection{Quadrant-covariance integrals}\label{subsec:qci}
As described in \citet{tenneti_stochastic_2016}, the fluctuating acceleration-velocity covariance dictates the temporal evolution of granular temperature in homogeneous fluidization of elastic particles. For a given particle, the sign of the covariance dictates if the acceleration acts as a source (positive; $a^{\prime}-v^{\prime}$ quadrants 1 \& 3) or a sink (negative; $a^{\prime}-v^{\prime}$ quadrants 2 \& 4). Therefore, to quantify the sources and sinks produced by the acceleration Langevin model, the quadrant-conditioned covariances must be computed
\begin{align}
2\left \langle v^{\prime}a^{\prime} \right \rangle &\equiv S - \Gamma = 4\int_{0}^{\infty} \int_{0}^{\infty} v^{\prime}a^{\prime} \, \mathcal{N}\left(\bar{\bm{\Sigma}}^{-1}\right) \, {\rm d}a^{\prime} \, {\rm d} v^{\prime} - 4\int_{-\infty}^{0} \int_{0}^{\infty} v^{\prime}a^{\prime} \, \mathcal{N}\left(\bar{\bm{\Sigma}}^{-1}\right) \, {\rm d}a^{\prime} \, {\rm d} v^{\prime},
    \label{eq:sginteg}
\end{align}
where $\mathcal{N}\left(\bar{\bm{\Sigma}}^{-1}\right)$ is the derived normal probability distribution with covariance matrix given in Eq.~\eqref{eq:covmat}, $S$ denotes the source, and $\Gamma$ denotes the sink. To evaluate the integrals given in Eq.~\eqref{eq:sginteg}, we proceed similar to the derivation provided in \citet{stuart_kendalls_2010}. For the source integral, substituting the Fourier transform of the characteristic function in place of the normal distribution function yields
\begin{align}
S = \frac{4}{(2\pi)^2} \int_{0}^{\infty} \int_{0}^{\infty} x_1 x_2  \int_{-\infty}^{\infty} \int_{-\infty}^{\infty} \exp\left(-i\bm{k}^{\intercal}\bm{x} - \frac{1}{2}\bm{k}^{\intercal} \bar{\bm{\Sigma}} \bm{k}\right) {\rm d}\bm{k} \,  {\rm d}\bm{x},
    \label{eq:q1integ}
\end{align}
where $\bm{x} = \left[v^{\prime} \, a^{\prime} \right]^{\intercal}$ and $\bm{k} = [k_{v^{\prime}} \, k_{a^{\prime}}]^{\intercal}$. Exchanging the order of integration and employing the identity
\begin{equation}
x_{j} \exp \left( -i k_j x_j \right) = i \diff{}{k_j} \left(\exp \left( -i k_j x_j \right)  \right),
    \label{eq:ftprod}
\end{equation}
where no summation over $j$ is implied, leads to
\begin{align}
S = \frac{1}{\pi^2} \int_{-\infty}^{\infty} \int_{-\infty}^{\infty} \exp\left(-\frac{1}{2}\bm{k}^{\intercal} \bar{\bm{\Sigma}} \bm{k}\right)  \prod_{j=1}^{2}\left( i \diff{}{k_j} \right) \int_{0}^{\infty} \int_{0}^{\infty} \exp\left(-i\bm{k}^{\intercal}\bm{x}\right) {\rm d}\bm{x} \, {\rm d}\bm{k}.
    \label{eq:q1integ2}
\end{align}
The inner most integrals over $\bm{x}$ may be interpreted as the Fourier transforms of the Heaviside function, which have principal values $\prod_j \left(i k_j \right)^{-1}$. After evaluating the derivatives, one obtains
\begin{align}
S = \frac{1}{\pi^2} \int_{-\infty}^{\infty} \int_{-\infty}^{\infty} \exp\left(-\frac{1}{2}\bm{k}^{\intercal} \bar{\bm{\Sigma}} \bm{k}\right)  \frac{1}{k_{v^{\prime}}^2 k_{a^{\prime}}^2}  {\rm d}\bm{k}.
    \label{eq:q1integ3}
\end{align}
The exponential in Eq.~\eqref{eq:q1integ3} will contain a covariance term $-\rho \sigma_{v^{\prime}} \sigma_{a^{\prime}} k_{v^{\prime}} k_{a^{\prime}}$, where $\rho$ is the correlation coefficient. Differentiating both sides twice with respect to $\rho$ yields
\begin{align}
\frac{{\rm d}^2S}{{\rm d}\rho^2} = \frac{\sigma_{v^{\prime}}^2 \sigma_{a^{\prime}}^2}{\pi^2} \int_{-\infty}^{\infty} \int_{-\infty}^{\infty} \exp\left(-\frac{1}{2}\bm{k}^{\intercal} \bar{\bm{\Sigma}} \bm{k}\right) {\rm d}\bm{k}.
    \label{eq:q1integ4}
\end{align}
The above integral will yield the reciprocal of the normalization constant for a bi-normal distribution
\begin{align}
\frac{{\rm d}^2S}{{\rm d}\rho^2} \equiv \frac{\sigma_{v^{\prime}}^2 \sigma_{a^{\prime}}^2}{\pi^2} 2 \pi \left| \bar{\bm{\Sigma}} \right|^{-1/2} = \frac{2\sigma_{v^{\prime}} \sigma_{a^{\prime}}}{\pi} \frac{1}{\sqrt{1 - \rho^2}}.
    \label{eq:q1integ5}
\end{align}
Integrating twice and determining the constants --- e.g. for sources $S=0; \, \rho = -1$ and $S=2\sigma_{v^{\prime}} \sigma_{a^{\prime}}; \, \rho = 1$ --- yields the following 
\begin{align}
S  = \frac{2\sigma_{v^{\prime}} \sigma_{a^{\prime}}}{\pi} \left(\rho \arcsin{\rho} + \sqrt{1-\rho^2} +  \frac{\pi}{2} \rho \right), \label{eq:sfinal} \\
\Gamma  = \frac{2\sigma_{v^{\prime}} \sigma_{a^{\prime}}}{\pi} \left(\rho \arcsin{\rho} + \sqrt{1-\rho^2} -  \frac{\pi}{2} \rho \right). \label{eq:gfinal}
\end{align}

\section*{Declaration of interests}
The authors report no conflict of interest.

\section*{Acknowledgements}
This material is based upon work supported by the National Science Foundation under grant no. CBET-1904742 and grant no. CBET-1438143.

\end{document}

%% file: macros.tex
\newcommand{\diff}[2]{\frac{\partial #1}{\partial #2}}

\newcommand{\diffn}[3]{\frac{\partial^{#3} #1}{{\partial #2}^{#3}}}

\newcommand{\odiff}[2]{\frac{{\rm d} #1}{{\rm d} #2}}